\documentclass[aps,preprint]{revtex4}%
\usepackage{amsfonts}
\usepackage{amsmath}
\usepackage{amssymb}
\usepackage{graphicx}%
\providecommand{\U}[1]{\protect\rule{.1in}{.1in}}

\begin{document}
\title[ ]{Magnetic Field Applied to the Classical Hydrogen Atom Treated in Classical
Electrodynamics with Classical Zero-Point Radiation}
\author{Timothy H. Boyer}
\affiliation{Department of Physics, City College of the City University of New York, New
York, New York 10031}
\keywords{}
\pacs{}

\begin{abstract}
An external magnetic field is applied to the classical hydrogen atom treated
in classical electromagnetic theory including classical electromagnetic
zero-point radiation. \ In an earlier article, it was shown that the average
value of each of the electron's action variables appears as a discrete value,
corresponding to a representation of the rotation group, because of
\textit{resonance }between the periodic orbit of the electron and the random
classical zero-point radiation. \ Here it is shown that, in the presence of a
magnetic field and because of resonance, the classical orbital motion of the
electron is in resonance with random classical zero-point radiation only for
orientations of the orbit which take integer values for the angle made with
the direction of the magnetic field, but excluding the $m=0$ orientation where
the magnetic field is parallel to the orbital plane of the electron.
\ Classical electromagnetic explanations are given for the Stern-Gerlach
result, and for the Zeeman effect.

\ 

\end{abstract}
\maketitle

\section{Introduction\ }

Recent work\cite{hydrogen2026} extending \textit{classical} electrodynamics
and including \textit{classical zero-point energy, relativity, and resonance},
has considered the hydrogen atom, but without imposing magnetic fields. \ In
this present article, we consider the effects of magnetic fields within the
same \textit{classical} electromagnetic theory. \ We will give
\textit{classical} explanations for aspects of the Zeeman effect, for the
Stern-Gerlach deflection, and for the Sommerfeld fine structure. \ The work
reported here involves only \textit{classical} electrodynamics and has no
concepts outside of \textit{classical} electrodynamics. \ 

The \textit{classical} electromagnetic analysis presented here presents such
an extremely different point of view from that of \textit{modern quantum}
theory that we believe it is necessary to assure the reader that only
\textit{classical} electromagnetic theory is involved. \ There are no ideas
which lie outside of \textit{classical} electrodynamics. \ However, there are
three \textit{unfamiliar} aspects considered.

1.In the first place, random \textit{classical} electromagnetic zero-point
radiation is taken as the experimentally observed fundamental distribution of
radiation inside of which all other aspects of classical electrodynamics take
place. \ The Casimir forces\cite{Casimir} between uncharged conduction plates
provide the quantitative experimental justification for the zero-point
radiation. \ A \textit{Lorentz invariant} distribution of random
\textit{classical} radiation with \textit{one} appropriate scaling factor will
account for both the observed spatial dependence and also the magnitude of
Casimir forces.\cite{B1973}

2.The restriction of allowed forces to only those allowed in
\textit{relativistic} classical electrodynamics is another unfamiliar aspect.
\ The Coulomb potential is part of \textit{relativistic} electrodynamics, as
are magnetic fields. \ The harmonic oscillator in one dimension appears when
one confines a charge (unstably) between two charges of the same sign and
takes the mass of the charge as very large.\cite{SHO}

3.Finally, the idea of \textit{resonance} for charged particles in random
classical electromagnetic zero-point radiation is a third unfamiliar aspect.
\ It is this resonant aspect which connects the action variables of the
present \textit{classical} electromagnetic theory with the \textit{quantum}
aspects of the \textit{old quantum} theory of Bohr, Wilson, and Sommerfeld.
\ The \textit{average} resonant action variable of the present
\textit{classical} theory agrees with the \textit{fixed} \textit{quantum}
aspect of the action variables of \textit{old quantum} theory. \ The
connection has been developed for circular orbits in previous
articles,\cite{hydrogen2026}\cite{SHO} and the addition of a magnetic field is
considered in the present article.

\section{Magnetic Field Applied to the Ground State of Hydrogen}

\subsection{Nonrelativistic Classical Hydrogen Atom}

The \textit{classical} model for a hydrogen atom consists of a very heavy
nucleus around which orbits an electron. \ In previous articles,\cite{B1975}%
\cite{hydrogen2026} we showed that the classical hydrogen atom has a lowest
energy state in the presence of random classical electromagnetic zero-point
radiation. \ Thus, in 1975, it was pointed out that at small orbital radii
where the electron is moving rapidly, the electron would absorb large amounts
of energy from the high frequency zero-point radiation; at large radii here
the electron is moving slowly, it would not gain enough energy from
low-frequency zero-point radiation to balance the loss of energy through
radiation emission. \ Therefore, for some radius in between the extremes,
there must be a balance between the energy loss due to radiation emission and
the energy gain from zero-point radiation. \ This average radius should serve
as the equilibrium radius of the classical hydrogen atom. \ However, in 1975,
there was no suggestion either of the importance of \textit{relativity} nor of
any \textit{resonant} aspect.\cite{B1975} \ 

In recent work,\cite{hydrogen2026} both relativity and the resonant aspect are
emphasized, but there is no magnetic field in the analysis. \ In the ground
state, the motion of the electron is approximately a circular
orbit.\cite{Jackson} \ However, the orientation of the circle will change
slowly due to the random but isotropic nature of the classical zero-point
radiation. \ \ Accordingly, in the absence of any magnetic field, the
probability distribution for the electron is isotropic in space. \ 

\subsection{Classical Hydrogen Atom in the Presence of a Magnetic Field}

In the presence of any small magnetic field, the situation changes sharply.
\ Consider first the situation when the magnetic field is parallel to the
circular orbit of the electron. \ If the orientation of the circular orbit is
taken as the $xz$-plane and the magnetic field is parallel to the $z$-axis,
then the classical magnetic force $e\mathbf{v\times B/}c~$will move the
particle out of the $xz$-plane. \ The mechanical motion of the charge
corresponds to a cork-screw, spiral motion corresponding to precession of the
circular orbit around the $z$-axis. \ In classical electromagnetic theory, the
charge is accelerating due to both the Coulomb force and the magnetic force,
and therefore the charge loses energy due to radiation emission. \ In order to
maintain the orbital precession, the charge must absorb energy from the random
classical zero-point radiation. \ As explained in earlier
work,\cite{hydrogen2026} the motion must repeat approximately the same orbit
in order to absorb the energy and balance the loss of energy due to radiation
emission on acceleration. \ Such repeated motion, where the electron energy is
balanced, represents precisely the \textit{resonance} between the random
classical zero-point radiation and the electron's orbital motion. \ However,
for the orientation of the electron's circular orbit parallel to the field
$\mathbf{B}$, the precession of the circular orbit prevents this repeated
energy gain. \ Thus the charge losses net energy and falls toward the nucleus,
experiencing random forces, until its orbit is approximately perpendicular to
the magnetic field. \ At this point when the motion is circular and the
magnetic field $\mathbf{B}$ is approximately perpendicular to the orbit, the
\textit{resonant} orbit remains a circular orbit, but now corresponding to
different frequencies depending upon the direction of the orbital motion
compared to the magnetic field. \ The force due to the Coulomb potential and
the force due to the magnetic field are now oriented so as to either add or
subtract from each other. \ 

For small magnetic fields $B,$ the \textit{radius} $r_{0}$ of the equilibrium
circular orbit remains unchanged by the presence of a small magnetic
field.\cite{Griffiths271} \ The nonrelativistic radial equation of motion for
the circular electron orbit with the magnetic field perpendicular to the
orbital plane is given by
\begin{equation}
M_{0}\omega^{2}r=\frac{e^{2}}{r_{0}^{2}}\pm\frac{e\omega r}{c}B.
\end{equation}
If the frequency in the absence of the magnetic field is $\omega_{0}$ from
\begin{equation}
M_{0}\omega_{0}^{2}r=\frac{e^{2}}{r_{0}^{2}},
\end{equation}
then we find%
\begin{equation}
\omega_{\pm}^{2}=\omega_{0}^{2}\pm\frac{e\omega B}{M_{0}c}\text{ \ \ or
\ \ }\omega_{\pm}\approxeq\omega_{0}\pm\frac{eB}{2M_{0}c}. \label{omegapm}%
\end{equation}

\subsection{No \textquotedblleft Spin\textquotedblright\ for the
\textit{Classical} Electron}

It is important to understand the \textit{classical} logic in the previous
paragraphs since it is so different from that offered in \textit{modern
quantum} texts. \ In the \textit{classical} \textit{mechanical} analysis in
the absence of a magnetic field, the ground state of the hydrogen atom is
moving in a circular orbit with nonrelativistic energy $\mathcal{E}$ given
by\cite{Goldstein476} \
\begin{equation}
\mathcal{E}=-\frac{M_{0}e^{4}}{2\left(  J_{r}+J_{2}\right)  ^{2}}=-\frac{1}%
{2}M_{0}c^{2}\left[  \frac{e^{2}}{\left(  J_{r}+J_{2}\right)  c}\right]  ^{2}.
\end{equation}
The moment that a magnetic field $\mathbf{B}$ (however small) is present,
there is a preferred orientation and also a nonrelativistic kinetic energy
change. \ For a small magnetic field causing a small change $\Delta v$ in the
electron's speed $v_{0}$, the nonrelativistic \textit{resonant} kinetic energy
of the electron due to the magnetic field is given by%
\begin{align}
KE &  =\frac{M_{0}}{2}\left(  v_{0}\pm\Delta v\right)  ^{2}\approxeq
\frac{M_{0}}{2}v_{0}^{2}+\frac{1}{2}M_{0}\left[  \pm2v_{0}\left(  \Delta
v\right)  \right]  \nonumber\\
&  =\frac{1}{2}M_{0}v_{0}^{2}\pm\left(  Mv_{0}r_{0}\right)  \frac{\Delta
v}{r_{0}}=\frac{1}{2}M_{0}v_{0}^{2}\pm\left\vert L_{z}\right\vert \frac
{eB}{2M_{0}c},
\end{align}
from Eq. (\ref{omegapm}), while the Coulomb potential energy is unchanged
since the radius $r_{0}$ is unchanged. \ Thus the energy of the ground state
of the classical hydrogen at is now split into two separate levels. \ The
orbital angular momentum $L_{z}$ of the original circular orbit with no
magnetic field is $L_{z}=Mv_{0}r$. \ In the presence of the small magnetic
field, the energy goes up by $\left\vert L_{z}\right\vert eB/\left(
2M_{0}c\right)  $ or goes down by $-\left\vert L_{z}\right\vert eB/\left(
2M_{0}c\right)  $ depending on the direction of the orbital motion, whether
clockwise or counter clockwise compared to the orientation of the magnetic
field and depending on whether the magnetic field contribution adds or
subtracts from the previous orbital motion. \ Thus the difference between
these two energies is just
\begin{equation}
\Delta\mathcal{E}=2\left(  \frac{eL_{z}}{2M_{0}c}\right)  B=\left(
\frac{eL_{z}}{M_{0}c}\right)  B.\ \label{DU}%
\end{equation}

On the other hand, the change of angular momentum for each of the orbital
shifts is
\begin{equation}
\Delta L_{z}=r^{2}M\,_{0}\Delta\omega=\pm r_{0}^{2}M_{0}\frac{eB}{2M_{0}c}.
\end{equation}
Note the factor of $2$ difference in the angular momentum shift and double
that shift for the energy difference between the two orbital motions. \ 

In \textit{classical} theory, the shift by this factor of $2$ arises from
\textit{classical} mechanical aspects. \ However, in \textit{modern quantum}
theory, the electron wave function in the absence of a magnetic field is
regarded as isotropic, since there is no discussion of the electron's orbital
motion. \ The factor of $\ 2$ for the energy difference in Eq, (\ref{DU}) is
usually described as arising from an intrinsic electron \textquotedblleft
spin.\textquotedblright\ \ The \textit{quantum} \textquotedblleft
spin\textquotedblright\ angular momentum $\mathbf{S}$ for the electron is half
as large as the orbital angular momentum and has eigenvalues
\begin{equation}
s=\pm\hbar/2.
\end{equation}
\ On the other hand, the magnetic moment of the electron is twice as large as
the angular momentum associated with the orbital motion\cite{Griffiths277279}%
\begin{equation}
\overrightarrow{\mu}_{S}=\frac{e}{M_{0}c}\mathbf{S.}%
\end{equation}
\ 

The \textit{quantum} description is completely different from that of the
\textit{classical} hydrogen description given above. \ However, one should
emphasize that the \textit{classical} electromagnetic analysis must have
\textit{resonance }with classical zero-point radiation\textit{ }as an
important aspect of the electron's orbital behavior. \ The resonant situation
(between the charge's motion and the random zero-point radiation) limits the
average behavior to that near the average balance situation for both the
orbital energy and angular momentum in the electron ground state. \ This
balance point gives average values to the \textit{classical} action variable
which agree with the discrete action variables assumed by \textit{old quantum}
theory. \ The \textit{classical} analysis does not agree with the descriptions
given in \textit{modern quantum} theory which starts with a
spherically-symmetric ground state and has no orbits. \ In the
\textit{classical} analysis, there is no electron spin, but there are rather
two \textit{resonant} classical particle trajectories leading to a two-way
splitting of the ground state due to the magnetic field. \ 

\subsection{Stern-Gerlach Experiment}

The \textit{resonant} classical situation also leads to a natural
\textit{classical} explanation of the Stern-Gerlach experiment. \ In the
Stern-Gerlach experiment\cite{SG} of 1922, it was found that a beam of silver
atoms was deflected into two distinct, separate paths to the collection plate.
\ Later, in 1927, the experiment was repeated with hydrogen atoms by Phipps
and Taylor.\cite{PT} \ Again, at low temperatures, exactly \textit{two}
distinct paths to the collection plate were found. \ In a \textit{classical}
electromagnetic calculation which did \textit{not} include zero-point
radiation and the possibility of resonance between the orbital motion and
zero-point radiation, such behavior was regarded as
surprising.\cite{Griffiths181183} \ A smooth continuous distribution was
suggested as the classical \textit{mechanical} result. \ 

In both experiments of the 1920s, the atoms have a charge valence of $+1$.
\ \ Therefore, we believe that there is essentially one outer electron which
can interact with a magnetic field. \ The external magnetic field arranged by
the experimental apparatus provided a preferred direction perpendicular to the
trajectory of the atomic beam. \ Thus when classical zero-point radiation is
included, the orbital motion was split between those resonant electron
trajectories in a clockwise and those in a counter clockwise direction
relative to the magnetic field direction. \ Each charge particle orbit in the
ground states would have had exactly \textit{two} possible directions of
\textit{resonant} revolution, and hence exactly two distinct paths of
deflection on the way to the collection plate. \ The electrons are moving with
frequencies of the order of $10^{15}$ revolutions per second would have had
plenty of time to assume a resonant configuration. \ 

This experiment is often cited as something which classical theory could not
possibly explain. \ However, these claims never include the idea of
\textit{classical} electromagnetic zero-point radiation nor the idea of
\textit{resonance} between the orbital motion of the charged particle and the
random classical zero-point radiation. \ 

\subsection{Specific Angular Momentum Values}

The \textit{classical} electromagnetic analysis for hydrogen involves no
\textquotedblleft spin\textquotedblright\ angular momentum. \ It also involves
changes in the ideas of angular momentum compared to \textit{modern quantum}
theory. \ \ Thus the ground state of hydrogen in \textit{modern quantum}
theory is taken as isotropic in its angular spatial distribution. \ In
contrast, the \textit{classical} ground state (lowest energy level for the
electron when zero-point radiation is included) is isotropic in space as long
as no magnetic field is present. \ In any magnetic field, the
\textit{resonant} orbital motion tends to orient itself in circular orbits
perpendicular to the magnetic field direction. \ Thus the ground state, in the
presence of any magnetic field, has resonant angular momentum values
$L_{z}=\pm\hbar$ depending on the direction of orbital rotation. \ 

The \textit{classical resonant} result $L_{z}=\pm\hbar$ is \textit{not} an
isotropic $l=0$ state of \textit{modern quantum} theory, nor of group theory.
\ Thus the \textit{classical} viewpoint would suggest a different set of
angular momentum values from those familiar from \textit{modern}
\textit{quantum} mechanics based upon solutions of the Schroedinger equation.
\ The \textit{modern} \textit{quantum} value $l=0$ suggests the identity
representation of the rotation group. \ On the other hand, the
\textit{resonant classical} viewpoint suggests a vector representation but
excludes the $m=0$ orientation. \ This orientation is excluded, not because
the electron would go through the Coulomb center, but rather because it would
not be \textit{resonant} between the orbital motion and classical zero-point
radiation. \ 

\subsection{\textquotedblleft Two-Valuedness\textquotedblright\ and
\textquotedblleft Space Quantization\textquotedblright}

If a \textit{classical} particle orbit is circular, it corresponds to an
action value $J_{r}=0$. \ Because of the \textit{resonance} between the
orbital motion and the zero-point radiation, the circular orbit with total
angular momentum\cite{Goldstein476} $J_{2}=J_{\theta}+J_{\phi}$ can assume
only certain orientations relative to the preferred $z$-axis; these
orientations correspond to the angular momentum along the $\widehat{z}$-axis
given by $J_{\phi}/\hbar=m=-l,-l+1,...,-2,-1,+1,+2,...,+l-1,+l$. \ Note that
the value $J_{\phi}/\hbar=m=0$ is excluded because the orientation would not
allow \textit{resonance} between the orbital motion and zero-point radiation.
\ This exclusion would hold for all values of the orbital angular momentum.
\ Thus, always, there is a set of positive values of $m$ and a separate set of
negative values of $m$. \ Thus always there is a two-valuedness associated
with the direction of motion of the orbit around the $z$-axis, whether
clockwise or counter clockwise.

This situation of discrete values for the resonant \textit{classical} angle
made with the preferred $z$-axis looks analogous to the \textquotedblleft
space quantization\textquotedblright\ of \textit{old quantum} theory. \ Thus,
the discrete set of values for $m$ suggests a discrete set of values of the
angle of orientation of the orbital motion in connection with the $z$-axis
which is the direction of the magnetic field. \ In the ground state, there are
two orientations corresponding to the \textit{resonant} \textit{classical}
values $L_{z}=\pm\hbar.$ \ In the lowest excited state, the \textit{resonant
classical} values for a circular orbit would be $L_{z}/\hbar=-2,-1,+1,+2.$
\ Note that the value $m=0$ is excluded as nonresonant. \ In the ground state,
the motion of the orbiting electron is in phase with the local average of the
random radiation. \ For an excited state, the radiation is at a higher
frequency than the orbital electron motion, and the average zero-point
radiation passes over the the orbiting electron twice. \ As shown in recent
work,\cite{hydrogen2026} each time the radiation field passes over the
orbiting charge, the charge picks up the same amount of energy. \ Thus, even
though the value of $m$ may be different, the radiation frequency at the
highest relevant frequency determines the amount of energy that the orbiting
charge receives. \ 

\section{Zeeman Effect for Hydrogen}

\subsection{Fine Structure of Hydrogen}

If a magnetic field is present when an electron changes from one resonant
excited state to some other state, the electron's energy change will involve
the difference between the energy of the initial resonant excited state and
the final state. \ Such energy levels will depend on the fine structure of the
hydrogen energy levels. \ A \textit{classical} approach to the fine structure
leads to the Zeeman effect in hydrogen and also in alkali metals involving
ideas very different ideas from those in \textit{modern quantum} theory. \ 

The fine structure calculated for the hydrogen energy levels agrees with the
result obtained by Sommerfeld\cite{Sommerfeld} in \textit{old quantum} theory,
but now reinterpreted \textit{classically} in terms of resonant behavior for a
charged particle in \textit{classical} zero-point radiation. \ We will work
with the classical \textit{relativistic} expression given in Goldstein's text
on \textit{classical} mechanics\cite{GoldsteinS}%
\begin{equation}
U(J_{3},J_{2,}J_{1})=M_{0}c^{2}\left[  1+\left(  \frac{e^{2}/c}{J_{3}%
-J_{2}+\sqrt{J_{2}^{2}-\left[  e^{2}/c\right]  ^{2}}}\right)  ^{2}\right]
^{-1/2}, \label{UJJJ}%
\end{equation}
where $J_{1}=J_{\phi},J_{2}=J_{\theta}+J_{\phi},~and~J_{3}=J_{r}+J_{\theta
}+J_{\phi}$. \ Expanding in $e^{2}/\left[  J_{3}c\right]  $ but suppressing
the $e^{2}/c$, we find the approximation%
\begin{align}
&  U(J_{3},J_{2,}J_{1})=M_{0}c^{2}\left[  1+\frac{1}{J_{3}^{2}}\left(
\frac{1}{1-\left(  J_{2}/J_{3}\right)  +\sqrt{\left(  J_{2}/J_{3}\right)
^{2}-1/J_{3}^{2}}}\right)  ^{2}\right]  ^{-1/2}\nonumber\\
&  =M_{0}c^{2}\left[  1-\frac{1}{2J_{3}^{2}}\left\{  \left(  1+\frac{1}%
{J_{3}J_{2}}+...\right)  +\right\}  +\frac{3}{8}\frac{1}{J_{3}^{4}}+..\right]
\end{align}
or, restoring the $e^{2}/c,$%
\begin{equation}
U=M_{0}c^{2}-\frac{M_{o}c^{2}}{2}\left(  \frac{e^{2}}{J_{3}c}\right)
^{2}-\frac{M_{o}c^{2}}{2}\left(  \frac{e^{2}}{J_{3}c}\right)  ^{2}\frac{J_{2}%
}{J_{3}}\left(  \frac{e^{2}}{J_{2}c}\right)  ^{2}+\frac{3}{8}M_{0}c^{2}\left(
\frac{e^{2}}{J_{3}c}\right)  ^{2}\left(  \frac{e^{2}}{J_{3}c}\right)  ^{2}
\label{ClassFS}%
\end{equation}
The approximate result in Eq. (\ref{ClassFS}) is that obtained within
\textit{classical} mechanics for a particle in a Coulomb (or Kepler)
potential. \ 

As was shown in the earlier article,\cite{hydrogen2026} the excited states of
hydrogen correspond to resonance between the particle orbital motion and
classical electromagnetic zero-point radiation. \ If we write, $J_{3}%
=n_{3}\hbar$ and $J_{2}=n_{2}\hbar,$ then the approximation in Eq.
(\ref{ClassFS}) becomes%
\begin{equation}
U\left(  n_{3},n_{2},n_{1}\right)  =M_{0}c^{2}-\frac{M_{o}c^{2}}{2n_{3}^{2}%
}\left(  \frac{e^{2}}{\hbar c}\right)  ^{2}-\left[  \frac{M_{o}c^{2}}%
{2n_{3}^{2}}\left(  \frac{e^{2}}{\hbar c}\right)  ^{4}\right]  \left[
\frac{1}{n_{3}n_{2}}-\frac{3}{4n_{3}^{2}}\right]  . \label{SommEq}%
\end{equation}
The first term in Eq. (\ref{SommEq}) is the relativistic \textit{rest energy}
of the charged particle, the second term is the \textit{Bohr energy}, and the
third term is the \textit{fine-structure energy} of hydrogen. \ As is often
mentioned in the physics literature, the Sommerfeld result (\ref{UJJJ}) in
\textit{old quantum} theory agrees with that of the Dirac
equation.\cite{Griffiths274}

\subsection{Zeeman Effect for the Ground State}

In the ground state, the orbital motion of the electron is expected to be
circular, since this is the most stable orbital configuration for the electron
in a Coulomb potential in the presence of random classical
radiation.\cite{Jackson} \ The orbital angular momentum is $L_{z}=\pm\hbar$,
and the average orbital \textit{nonrelativistic} energy is the Bohr ground
state energy from Eq. (\ref{SommEq}) plus the magnetic from Eq. (\ref{DU}),
\begin{equation}
\mathcal{E=}-\frac{M_{0}e^{4}}{2\hbar^{2}}\pm\hbar\frac{eB}{2M_{0}c}.
\end{equation}
This expression agrees with Griffiths expression for the weak-field splitting
of the ground state of hydrogen.\cite{Griffiths279}

\subsection{Zeeman Effect for Bohr's $n=2$ Levels}

For Bohr's excited level $n=2$ in Eq. (\ref{ClassFS}), we have $J_{3}%
=2\hbar.~$However, now either we can have $J_{r}=\hbar$ and $J_{2}=\hbar,$ or
we can have $J_{r}=0$ and $J_{2}=2\hbar$. \ The eccentricity of the
\textit{nonrelativistic resonant} classical orbit is given by%
\begin{equation}
\epsilon=\sqrt{1-\left(  \frac{J_{2}}{J_{3}}\right)  ^{2}}.
\end{equation}
Thus the eccentricities are different for these orbits since the ratios of
$J_{2}/J_{3}$ are different. \ The lower value of $J_{2},$ where $J_{r}=1$
corresponds to nonrelativistic elliptical (relativistic rosette) orbits with
smaller angular momentum where the nonrelativistic eccentricity is $\sqrt
{3/4}.$ \ On the other hand, the orbits where $J_{3}=J_{2}$, $J_{r}=0,$ have
eccentricity zero, corresponding to a \textit{circular} orbit since $J_{r}=0$.
\ In the absence of a magnetic field, the orbit can have any orientation. \ If
a magnetic field is present, the \textit{classical resonant} orbit will be
precessing around the magnetic field direction but $m=0$ is excluded.

We see from the $1/\left(  n_{3}n_{2}\right)  $ dependence in Eq.
(\ref{SommEq}) that the levels where the resonant value of $n_{2}=l$ is
smaller have lower (deeper) energy levels compared to those where $n_{2}=l$ is
larger. \ Thus we expect that the the energies for $n_{3}=2,l=1$ are lower
than those for $n_{3}=2,l=2.$ \ There are two lower energy levels,
corresponding to $n_{3}=1,l=1,m=\pm1$,~and these will exhibit the same doublet
sort of splitting in a magnetic field as found for the ground state, which was
discussed above. \ On the other hand, the higher levels, corresponding to
$n_{3}=2,l=2,m=\pm2,\pm1,$ will exhibit a four-way splitting in a weak
magnetic field. \ Remember that the orientation $m=0$ is excluded as not
resonant by the \textit{classical resonant} analysis. \ The four-way splitting
seems similar to Griffiths diagram.\cite{Griffiths282}

In our \textit{classical} electromagnetic analysis, the \textit{electron is
simply a point charge}, and there is no such thing as\textit{ }a\textit{
\textquotedblleft spin-orbit\textquotedblright\ interaction}. \ In contrast,
\textit{modern quantum} theory regards the fine structure as arising from two
\textit{separate} effects, the relativistic correction and the spin-orbit
coupling. \ In his \textit{quantum} text, Griffiths notes\cite{Griffiths274}
\textquotedblleft It is remarkable, considering the totally different physical
mechanisms involved, that the relativistic correction and the spin-orbit
coupling are of the same order $(\mathcal{E}_{n}^{2}/Mc^{2}).$%
\textquotedblright\ \ In the \textit{classical} view, the situation is not
remarkable at all and has nothing to do with \textquotedblleft spin-orbit
coupling;\textquotedblright\ it is simply the result of \textit{classical}
electrodynamics. \ \ We notice from the original relativistic expression, Eq.
(\ref{UJJJ}), that $J_{2}$ cannot vanish since it appears in a square root
with a subtraction.

\section{Closing Comments}

In this article, we extend the earlier work\cite{B1975} on \textit{classical}
electrodynamics which includes classical electromagnetic zero-point radiation.
\ The analysis is entirely \textit{classical} electromagnetic. \ It appears
that some more of the aspects of atomic physics can be explained
\textit{classically} without the need to invoke \textit{quantum} notions.
\ The Zeeman effect for hydrogen can be explained in terms of
\textit{classical} electrodynamics without the need for \textquotedblleft
electron spin.\textquotedblright\ \ \ Indeed, at least for the lower levels of
hydrogen, \textquotedblleft spin\textquotedblright\ seems to refer to the
direction of motion of an electron within its classical orbit. \ The
direction, whether clockwise or counter clockwise, provides a two-valuedness
for both hydrogen and also for one-outer-electron atoms.

We must be sure that the orbital motion will allow \textit{resonance} with
classical zero-point radiation. \ The Sommerfeld relativistic result for
hydrogen can be understood as arising from \textit{classical} theory where the
action variables of old quantum theory take integer values because of
\textit{resonance} between the orbital motion and zero-point radiation.
\ \textit{Classical resonance} explains the \textquotedblleft space
quantization\textquotedblright\ of old quantum theory. \ Furthermore, the
Stern-Gerlach experiment can be explained within \textit{classical}
electromagnetic theory.

\section{Acknowledgement}

I wish to thank Professor Daniel C. Cole for helpful comments on an earlier
version of this manuscript.


\begin{thebibliography}{99}                                                                                               %


\bibitem {hydrogen2026}T. H. Boyer, \textquotedblleft Relativistic Hydrogen in
Classical Electrodynamics with Classical Zero-point
Radiation,\textquotedblright\ submitted for publication., arXiv 2603.13448.

\bibitem {Casimir}H. B. G. Casimir, \textquotedblleft On the attraction
between two perfectly conducting plates,\textquotedblright\ Proc. Ned. Akad.
Wetenschap. \textbf{51}, 793-795 (1948).

\bibitem {B1973}T. H. Boyer, \textquotedblleft Retarded van der Waals Forces
at All Distances Derived from Classical Electrodynamics with Classical
Electromagnetic Zero-Point Radiation,\textquotedblright\ Phys. Rev. A
\textbf{7}, 1832-1840 (1973).

\bibitem {SHO}T. H. Boyer, \textquotedblleft The Classical Linear Oscillator
in Classical Electrodynamics with Classical Zero-Point
Radiation,\textquotedblright\ submitted for publication, arXiv 2603.13446.

\bibitem {B1975}T. H. Boyer, \textquotedblleft Random electrodynamics: The
theory of classical electrodynamics with classical electromagnetic zero-point
radiation,\textquotedblright\ Phys. Rev. D \textbf{11}, 790-808 (1975).

\bibitem {Jackson}J. D. Jackson, \textit{Classical Electrodynamics 2nd ed}
(John Wiley \& Sons, New York, 1975), p. 784.

\bibitem {Griffiths271}D. J. Griffiths, \textit{Introduction to
Electrodynamics} \textit{5th ed} (Cambridge U. Press, Cambridge 2024), pp. 273-274.

\bibitem {Goldstein476}H. Goldstein, \textit{Classical Mechanics 2nd ed},
(Addison-Wesley, Reading, MA 1981), 476.

\bibitem {Griffiths277279}See D. J. Griffiths, \textit{Introduction to Quantum
Mechanics 2nd ed} (Pearson Prentice Hall, Upper Saddle River, NJ 2005), pp. 277-279.

\bibitem {SG}W. Gerlach and O. Stern, \textquotedblleft Der experimentelle
Nachweis der Richtungsquantelung im Magnetfeld\textquotedblright\ Zeitschrift
f\"{u}r Physik \textbf{9}, 349--352 (1922). \ 

\bibitem {PT}T. E. Phipps and J. B. Taylor, \textquotedblleft The Magnetic
Moment of the Hydrogen Atom,\textquotedblright\ Physical Review \textbf{29},
309--320 (1927).

\bibitem {Griffiths181183}See ref. 9, pp.181-183. \ 

\bibitem {Sommerfeld}A. Sommerfeld, \textquotedblleft Zur Quantentheorie der
Spektrallinien,\textquotedblright\ Annalen der Physik, \textbf{356}, 1--94
(1916). \ 

\bibitem {GoldsteinS}See ref. 8, p. 498.

\bibitem {Griffiths274}See ref. 9 pp. 274-276.

\bibitem {Griffiths279}See ref. 9, p. 279.

\bibitem {Griffiths282}See ref. 9, p. 282.

June 7, 2026 \ \ \ \ \ \ \ \ \ \ \ MagneticField.tex
\end{thebibliography}
\end{document}